\documentclass{iau}
\usepackage{graphicx}
\usepackage{natbib}

\newcommand{\cv}{C$_{24}$\,}
\newcommand{\cs}{C$_{60}$\,}
\newcommand{\csp}{C$_{60}^+$}

\title[C$_{60}^+~in~space$] 
{Detection of the buckminsterfullerene cation (C$_{60}^+$) in space}

\author[Olivier Bern\'e \& Giacomo Mulas \& Christine Joblin]   
{Olivier Bern\'e$^{1,2}$, Giacomo Mulas$^3$
 \and Christine Joblin$^{1,2}$
 }

\affiliation{$^1$Universit\'e de Toulouse; UPS-OMP; IRAP;  Toulouse, France \\[\affilskip]
$^2$CNRS; IRAP; 9 Av. colonel Roche, BP 44346, F-31028 Toulouse cedex 4, France \\[\affilskip]
Istituto Nazionale di Astrofisica -- Osservatorio Astronomico di Cagliari -- strada 54, localitˆ Poggio dei Pini, 09012-- Capoterra (CA), Italy}

\pubyear{2013}
\volume{xxx}  
\pagerange{119--126}
\jname{Title of your IAU Symposium}
\editors{A.C. Editor, B.D. Editor \& C.E. Editor, eds.}
\begin{document}

\maketitle

\begin{abstract}
In the early 90's, C$_{60}^+$ was proposed as the carrier of two diffuse interstellar bands (DIBs) at 9577 and 9632\,\AA, but a firm identification still awaits gas-phase spectroscopic data. Neutral C$_{60}$, on the other hand, was recently detected through its infrared emission bands in the interstellar medium and evolved stars.  In this contribution, we present the detection of C$_{60}^+$  through its infrared vibrational bands in the NGC 7023 nebula, based on spectroscopic observations with the \emph{Spitzer} space telescope, quantum chemistry calculation, and laboratory data from the literature. This detection supports the idea that C$_{60}^+$ could be a DIB carrier, and provides robust evidence that fullerenes exist in the gas-phase in the interstellar medium. Modeling efforts  to design specific observations, combined with new gas-phase data, will be essential to confirm this proposal. A definitive attribution of the 9577 and 9632\,\AA\,DIBs to \csp\, would represent a significant step forward in the field. 
\end{abstract}

\firstsection 
\section{Introduction}

The mid-infrared (mid-IR) spectrum of galactic and extragalactic objects exhibits band emission (strongest at 3.3, 6.2, 7.7, 8.6, and 11.2 $\mu$m)
attributed to carbonaceous macromolecules, i.e., polycyclic aromatic hydrocarbons (PAHs, see recent state of the art in \cite{job11}). In addition to PAH bands, 
{ IR emission bands} at 7.0, 8.5, 17.4, and 19.0 $\mu$m have recently been reported \citep{cam10,sel10}, and found to 
{ match quite closely the IR active bands} of buckminsterfullerene (C$_{60}$,~\cite{kro85}), a cage-like carbon molecule. 
Carbonaceous macromolecules, {  including  PAHs}, carbon clusters, or fullerenes, are believed to play a fundamental role in the physics and chemistry
of the interstellar medium ({ISM)}, and their infrared signatures are commonly used as a tracer of physical conditions. Nevertheless, \cs is the only molecule belonging
 { to this family}, which has been specifically identified in the {  ISM}.
In the NGC 7023 reflection nebula, \cite{sel10} have shown that \cs\, is predominantly found in the regions closest to the star. In that part of the nebula,
{ UV} irradiation is high (above $10^4$ times the interstellar standard radiation field), and PAH molecules are ionized \citep{rap05, ber07, pil12},
{ if not destroyed \citep{ber12, mon13}. One could therefore expect \csp\, to be present in these regions.}

\cite{foi94} reported evidence for interstellar \csp\, based on the detection of { two} diffuse interstellar bands (DIBs) at 9577  and 9632 \AA,
{however this identification is still questioned considering that no spectroscopic information on  \csp\, could be recorded yet, in conditions appropriate for DIB
identification, i.e., in the gas-phase and at low temperature.}
{ The IR spectrum of \csp\, was measured in a rare-gas matrix by \cite{ful93} and was found to exhibit two bands at 7.1 and 7.5 $\mu$m.
 \cite{ker12} have performed new spectroscopic measurements and suggested that the latter band was due to C$_{60}^-$, whereas the authors
 attributed a band at 6.4 $\mu$m to \csp.} \cite{mou99} derived upper limits on the abundance of \csp\, in 
NGC 7023 based on ISO-SWS data. So far, there has been no observational evidence of any 7.1 or 6.4 $\mu$m bands in astronomical sources.

Looking carefully at the \emph{Spitzer} data of NGC 7023, we found four emission bands, 
at 6.4, 7.1, 8.2, and 10.5 $\mu$m, which are only present in the regions closest to the star. 
This also corresponds to a region where  \cs\, emission is strong. A natural carrier to explain 
these bands is \csp, and this assertion is { supported} by spectroscopic arguments that we 
discuss hereafter.

\section{Observational results} \label{obs_res}

\begin{figure*}
\begin{center}
\includegraphics[width=\hsize]{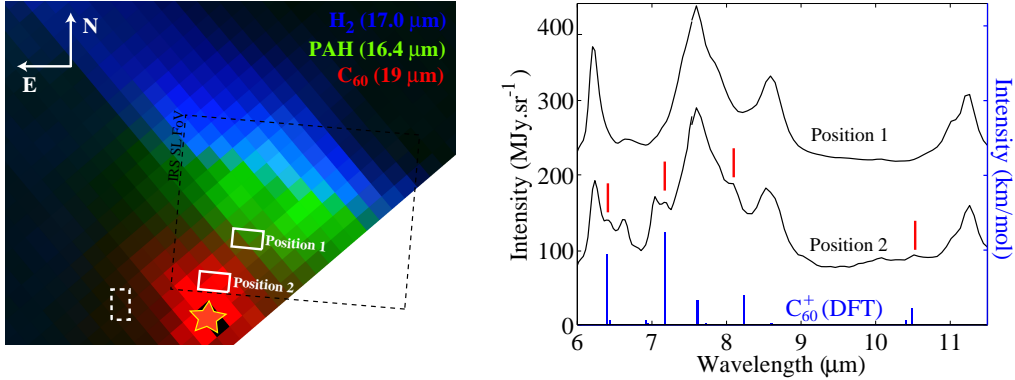}
\caption{ \emph{Left:} False-color image of the NGC 7023 nebula such as the one presented in \cite{sel10}, obtained from integrating
different components in the \emph{Spitzer}-IRS  LL spectral cube. Red is the emission integrated in the \cs 19 $\mu$m band.
Green is the emission of the PAH 16.4 $\mu$m band. Blue is the emission integrated in the H$_2$ (0-0) S(0) 17.0 $\mu$m 
band. The white rectangles indicate the regions where the IRS-SL spectra shown in the right panel have been extracted.
{ The black dashed rectangle depicts the IRS-SL field of view.} The white dashed rectangle shows the region where \cite{sel10} 
extracted their spectrum. \emph{Right:} Spectra at positions 1 and 2
in the image. The spectrum at position 1 has been shifted up and scaled down to allow easier comparison to the spectrum at position 2.
{ Error bars are not shown here but are comparable to the width of the line (see Fig. 2)}. The red lines indicate the four newly detected bands 
attributed to \csp. The DFT calculated spectrum (with scaled wavelength positions) is shown as a bar graph in blue.
\label{Spectra}}
\end{center}
\end{figure*}

Fig. 1 shows two mid-IR spectra extracted in NGC 7023 at two positions, one close the the star and one farther away.
These spectra have in common that they are dominated by bands at 6.2, 7.7, 8.6, and 11.2 $\mu$m, which are attributed to vibrational modes
of PAH molecules. In addition to these bands, the spectrum at 
position 2 shows several bands that are absent in the rest of the nebula. 
These are at $\sim$ 6.4, 6.6, 7.0, 7.1, 8.2, and 10.5 $\mu$m (Fig.~\ref{Spectra}).
{ These bands are seen in several pixels of the IRS cube and well above the 
instrumental errors \citep{ber13}}. To derive the precise parameters for these bands we fit them using Gaussian profiles  and spline
curves to subtract the underlying emission due to the wings of PAH bands. The positions, widths, and intensities of the bands are given in\citep{ber13}.
The 7.0 $\mu$m band has been attributed to \cs\, \citep{sel10}. The 6.6 $\mu$m band has recently been attributed to (possible) planar \cv \citep{gar11}.
 The 6.4, 7.1, 8.2, and 10.5 $\mu$m bands have not been observed or discussed yet. 
{ Since HD 200775 is a B star, only low ionization potential atoms should emit in fine structure lines (e.g. [CII], [SIII] etc.), and these
species do not have lines in this spectral range, so we exclude contamination by fine structure lines.}
These { four new bands seem spatially correlated, i.e. all of them only appear
in the regions closest to the star, which suggests a common carrier.} As shown by
 \cite{sel10}, \cs\, is also found only close to the star. { Still, the four new bands only appear in the regions that are
 closest to the star, while Fig. 1 demonstrates that \cs\, emission is more extended. This suggests that the four new bands
are carried by a species that is a product of the photoprocessing of \cs, an obvious carrier being \csp.}
In the following section we provide the spectroscopic arguments that support this observational evidence.

\section{Spectroscopic signatures of \csp}\label{spec}

{ The only IR spectroscopic data of  \csp\, have been obtained in rare gas matrices by \cite{ful93} and
more recently by \cite{ker12}. Two bands at 7.1 and  6.4\,$\mu$m seem definitively attributed
to \csp, based on these experimental studies.}
Theory is another approach to obtain spectroscopic data, and density functional 
theory (DFT) in particular has been shown to be effective and accurate for calculations 
on neutral \cs\, \citep{chase1992,fabian1996,iglesias-groth2011}.
However, an additional theoretical problem is that upon ionization C$_{60}$ is known to undergo 
Jahn\textendash Teller (JT) distortion \citep{cha97, ber06}. This is described in more details
in \cite{ber13} where we conclude that the accuracy of the calculated frequencies, guided by laboratory 
data, are nevertheless sufficient for an assignment. 
We performed our DFT calculations using the hybrid B3LYP exchange\textendash correlation functional 
and the 4\textendash 31g Gaussian basis set. This combination is known to yield reasonably accurate 
vibrational frequencies, after scaling with an empirical factor $\chi$ to account for the 
overestimation of the frequencies. 
In the case of PAHs, $\chi$ is calculated by comparing the computed frequencies to the ones measured in the laboratory at 
low temperature in rare gas matrices (see \cite{bau97} for a case-study on PAHs). Therefore, we used the  IR  absorption  
spectrum  of \csp\, that was measured in Ne matrices by Kern et al. (2012) to calibrate our DFT calculations. The two bands definitively 
attributed to \csp\, by \cite{ker12} are found at 1550 and 1406 cm$^{-1}$ and correspond to the strongest bands, predicted by theory at 1607 and 
1434 cm$^{-1}$. This implies a respective scaling  factor of 0.9645 and 0.9805. We adopt an average
value of $\chi=0.9725$. The resulting spectrum is shown in Fig. 1 (see \citealt{ber13} for the list of unscaled frequencies). 
After scaling, the five strongest IR bands fall at wavelengths of  6.40, 7.17, 7.60, 8.23 and 10.50 $\mu$m. 
Four of these are very close (within 2\%) to the positions of the four new  bands detected in NGC 7023 (6.43, 7.13, 8.10, 10.53 $\mu$m). 
The observed match is very good considering that other factors are expected to affect the band positions, in particular band shift 
due to anharmonic coupling in hot emitting molecules \citep{job95}.
The nondetection of the {7.6} $\mu$m band in the observations is not surprising, since it is most likely 
hindered by the strong PAH emission at the same position (Fig. 1).
Based on the observational (Sect. \ref{obs_res}) and spectroscopic arguments presented, 
we argue that there is strong evidence for the presence of \csp\, in NGC 7023.
Using the mid-IR intensity of the 4 bands detected in NGC 7023, we derive an abundance
of about 10$^{-4}$ of the elemental carbon to be locked in \csp (see \citealt{ber13} for details).

\section{Relation to DIBs}\label{sec_dibs}

The 9577 and 9632\,\AA~DIBs observed towards several hot stars have been attributed to the absorption of \csp~
present in the diffuse ISM \citep{foi94}. However, the laboratory spectroscopic data
on which this assignment relies were obtained in rare-gas matrices, and therefore a definitive attribution
still awaits gas phase laboratory data.  The detection presented here supports
the idea that \csp~is present in the ISM. However it should be noted that the abundances derived
from the mid-IR emission spectrum of NGC 7023 are a factor of at least 10 smaller than those found by \cite{foi94}
for the diffuse ISM. Since these environments are radically different, as well as the techniques used
to derive these abundances, one cannot argue that this discrepancy is significative. The large abundances 
found in the diffuse ISM could result from high yields of \cs and \csp~ formation by the photochemical
processing of PAHs under irradiation by massive stars, over large timescales \citep{ber12}. 
In any case, the rather low abundance of \csp~ found in NGC 7023 suggests that,
if they are indeed due to \csp, the 9577 and 9632\,\AA~DIBs are expected to have a very weak optical depth 
in this object. Detailed calculations to determine if these absorptions could be detected in NGC 7023 are ongoing, 
but as a first check we have looked for the presence of the 9577 DIB
in the near infrared spectrum of HD 200775. The data were obtained at the Pic du Midi 
observatory, with the NARVAL instrument, on the 2 meter Bernard Lyot telescope and with a resolution 
of 65 000 and presented in \cite{ale08}. The spectrum of HD 200775 was divided by a reference
star in order to remove the telluric lines which are ubiquitous in this spectral range. So far, we were unable 
to find any signature of the 9577\,\AA~DIB (Fig. 2, note that the 9632\,\AA~DIB falls outside of the wavelength 
 range observable with NARVAL). However, we insist that this non-detection 
cannot rule out the presence of the 9577\,\AA~DIB in NGC 7023, especially because the signal-to-noise
ratio obtained here is not optimal. Detailed modeling relying on the available laboratory data is needed to
determine if this band together with the 9632\,\AA~DIB are detectable. In the latter case, specific observations 
in the near-IR should be designed to maximize the chances of detection in the surroundings of HD 200775.
In this frame, new gas-phase spectroscopic data in this spectral range would be an invaluable asset.

\section{Conclusion}

After studying the mid-IR spectra of the NGC 7023 nebula, we have found spectral signatures 
at 6.4, 7.1, 8.1, and 10.5 $\mu$m, which we attribute to \cs\, in the cationic charge state (\csp). 
 \csp\, has been proposed as a DIB carrier, and our identification supports this proposal. 
Ultimately, one would want to identify \emph{in the same source} the spectroscopic signatures
of \csp, in emission and in absorption. Modeling efforts  to design specific observations,
combined with new gas-phase data will be essential to reach this goal which would represent
a significant step forward in the field of DIB research. Finally, we note that the detection presented
in this paper also provides direct evidence for the presence of large carbon molecules
in the gas phase in the ISM,  giving further support to the PAH hypothesis \citep{leg84}.

\begin{figure*}
\begin{center}
\includegraphics[width=\hsize]{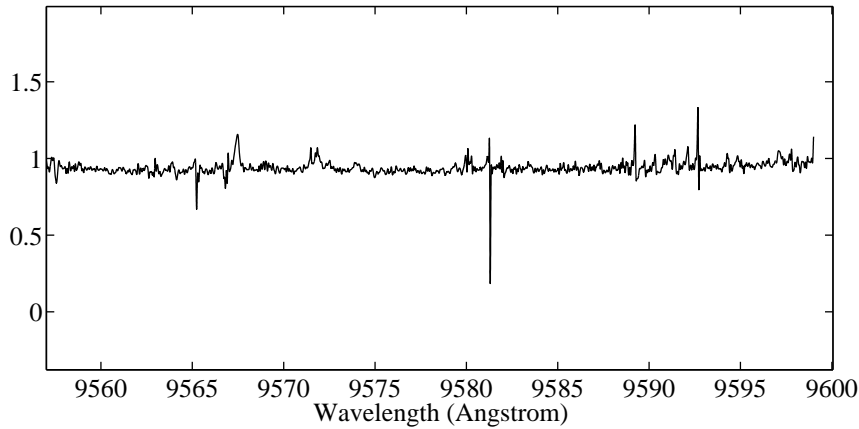}
\caption{Spectrum of HD 200775 obtained with the NARVAL 
spectrograph (zoom on the 9555-9600\,\AA~region). The absorption due to \csp\,
is expected to occur at 9577\,\AA~but is not seen with this level of signal-to-noise ratio.
The sharp peaks are residuals from telluric lines. Broad bands in emission are
stellar. 
\label{TBL}}
\end{center}
\end{figure*}

\bibliographystyle{aa}
\bibliography{biblio.bib}

\end{document}